\documentclass[aps,aps,twocolumn]{revtex4}

\usepackage{graphicx}
\usepackage{dcolumn}
\usepackage{bm}
\usepackage{amsmath}
\usepackage{txfonts}
\usepackage[scaled]{helvet}
\usepackage{color}
\usepackage[breaklinks=true,colorlinks,citecolor=blue,linkcolor=blue,urlcolor=blue]{hyperref}

\begin{document}

\title{Electrically Controllable van der Waals Antiferromagnetic Spin Valve}
\author{X.-C. Zhai$^{1,2}$, Ziming Xu$^3$, Qirui Cui$^{4}$, Yingmei Zhu$^{4}$, Hongxin Yang$^{4,5}$, Yaroslav M. Blanter$^2$}
\affiliation{$^1$Department of Applied Physics, Nanjing University of Science and
Technology, Nanjing 210094, China\\
$^2$Kavli Institute of NanoScience, Delft University of Technology,
2628 CJ Delft, The Netherlands\\
$^3$ School of Science, Nanjing University of Posts
and Telecommunications (NJUPT), Nanjing 210023, China\\
$^4$Ningbo Institute of Materials Technology and Engineering, Chinese
Academy of Sciences, Ningbo 315201, China\\
$^{5}$Center of Materials Science and Optoelectronics Engineering, University of Chinese Academy of Sciences, Being 100049, China}

\date{\today}

\begin{abstract}
We propose a spin valve that is based on van der Waals
antiferromagnetism and fully electrically controlled. The device
is composed of two antiferromagnetic terminals that allow for vertical bias control and a linked central scattering potential region. The magnetoresistance varies significantly when the bias orientations in two terminals are switched from parallel to antiparallel because this switch induces a mismatch of the bands for the same spin projection in different parts of the system. It is shown from density functional calculations that bilayer graphene encapsulated by two atomic layers of Cr$_2$Ge$_2$Te$_6$ provides a material platform to realize the antiferromagnetism, which is robust against the required vertical electric fields.
\end{abstract}

\maketitle

\section{Introduction}
Spin valve, consisting of two or more layered conducting ferromagnetic (FM) materials, is a key element of spintronics \cite{MonLod,ParkWund}.
The electrical resistance of a
spin valve switches between high and low values depending on
whether the alignment of the magnetization in the layers is parallel (P)
or antiparallel (AP). The mechanism of this change is the giant
magnetoresistance (MR) effect \cite{Grunberg}, which is the
basis of modern commercial devices such as reading heads for computer hard drives or magnetic sensors. In the past decade, spin valve also attracted much attention in 2D materials, see {\it e.g.} Refs.~[\onlinecite{YeZhu,ModMog,LuoM,LinYang,CarSor,SongCai,HanKaw0}], which have easily tunable electronic properties.

Recently-discovered 2D van der Waals (vdW) magnets, notably Cr$_2$Ge$_2$Te$_6$ (CGT) \cite{GongLi} and CrI$_3$ \cite{HuaCla}, have easily-manipulated magneto-electric and magneto-optical properties \cite{BurchMand,GoZha}, and therefore are very promising for spintronics. In particular, the layered antiferromagnetic (LAF, {\it i.e.} two adjacent FM monolayers coupled antiferromagnetically) order confirmed in bilayer insulating CrI$_3$ \cite{HuaCla,JiangLi,JiangShan,SunYi} adds an intriguing dimension to develop 2D antiferromagnetic spintronics with low power consumption and robustness against perturbations from magnetic fields \cite{BalMan,JungMar}. A few proposals exploiting LAF have been put forward, including manipulating electron spin by high-intensity electric field \cite{GongSJ} and exotic topological phases by electro-optical effect \cite{ZhaiBlanter}. However, the intrinsic LAF systems usually have large band gaps, no or weak spin polarization or low electron mobility \cite{HuaCla,BurchMand,GoZha,JiangLi,JiangShan,SunYi,GongSJ}, which strongly limit their applications in design of electronic circuits with low energy consumption.

Recent progress made it possible to fabricate 2D structures composed of almost any layers on top of each other \cite{LiYang,ZhongSey,NovMish}. These structures display a higher variety of properties than natural materials, can be tailored to have predefined physical properties, and open a way to design high-performance devices based on unusual interface physics. Motivated by the latest advances on fabrication of four-layer vdW heterosystems (typically graphene/bilayer CrI$_3$/graphene \cite{SongCai} and WSe$_2$/bilayer graphene (BLG)/WSe$_2$ \cite{Island}) and a BLG-based device that allows for local bias control \cite{LiWang}, we propose an artificial four-layer LAF device [Fig.~1(a)] with two gated terminals linked by the central scattering region to realize a fully electrically controlled spin valve. Its operation mechanism is similar to that of existing giant MR devices \cite{Grunberg,YeZhu,ModMog,LuoM,LinYang,CarSor,SongCai,HanKaw0},
with the difference that it does not require magnetic field: By altering the mutual bias orientation in the two terminals, from
P to AP, the MR of in-plane transport varies significantly. Due to the magneto-electric coupling [Fig.~1(b)], the spin valve effect is transferred to the nonmagnetic vdW layers. We start with
a phenomenological model, and further use the density functional theory
(DFT) calculations to demonstrate the concept with a material platform$-$BLG encapsulated by two single layers of CGT.

Our proposed LAF spin valve has three advantages over the usual spin valves \cite{Grunberg,YeZhu,ModMog,LuoM,LinYang,CarSor,SongCai,HanKaw0}: ($i$) It is robust against environmental perturbations from magnetic or
null-stray fields due to antiferromagnetism \cite{BalMan,JungMar,GongSJ};
($ii$) It requires only small vertical bias to control, is reversible to operate, superior to traditional magnetic control \cite{Grunberg}; ($iii$) It is not sensitive to the presence of domain walls (DWs), thus enabling the device to possess strong anti-jamming of DWs. The two latter points are not trivial, and we demonstrate their validity here.

\begin{figure}
\centerline{\includegraphics[width=8.5cm]{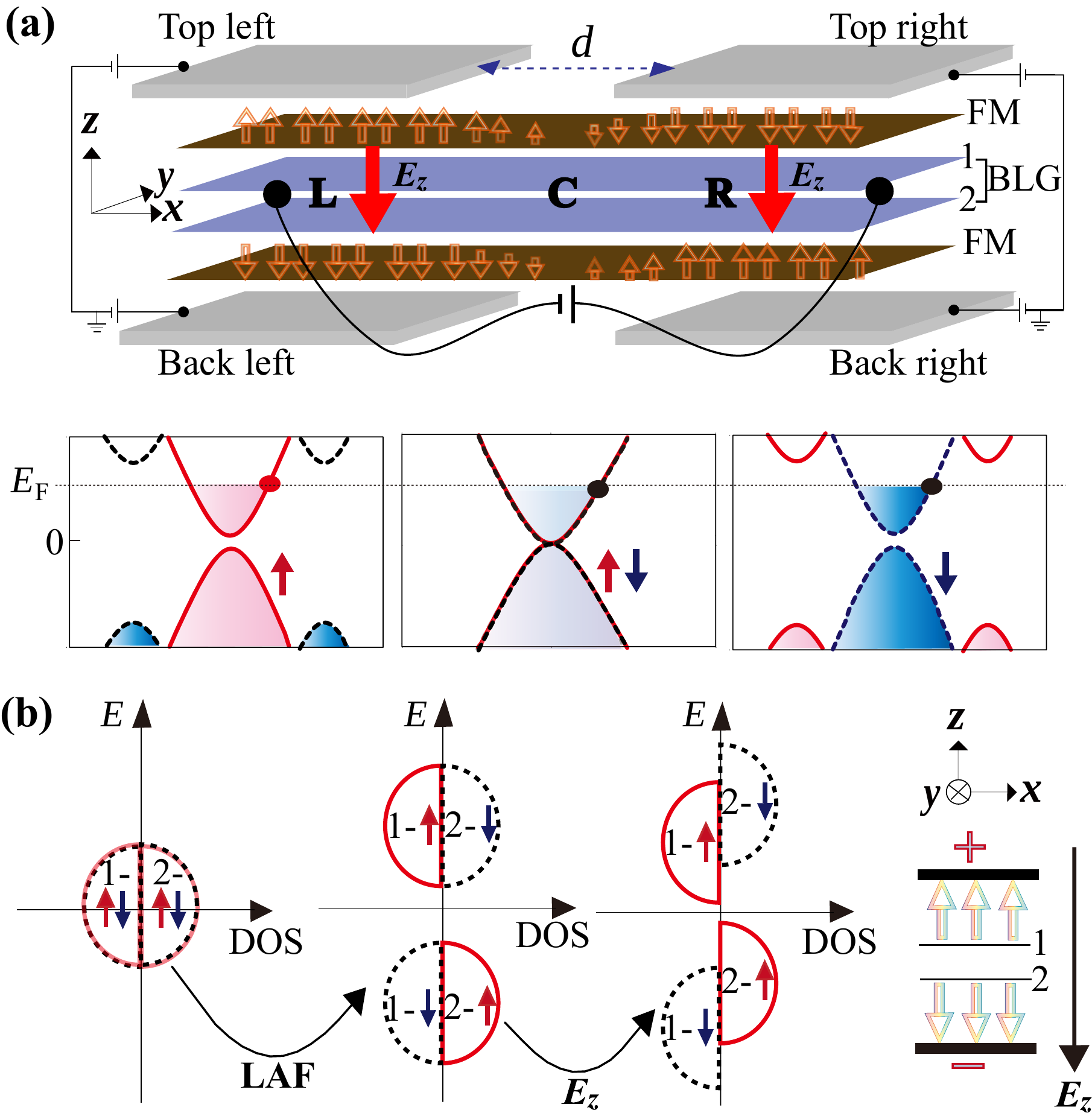}} \caption{
(a) Top: Sketch of a LAF spin valve based on BLG encapsulated by 2D FM insulators.
1 (2) marks the layer of BLG. The device consists of two LAF terminals [left (L), right (R)] and a linked central region (C, length $d$), where a collinear
DW (magnetic periodicity along $y$ is not drawn) is taken as an example. Local dual-gates in L/R are used to induce the vertical electric field $E_z$ and tune the Fermi energy $E_{\rm F}$ \cite{SongCai,LiWang}. Bottom: Typical bands are plotted to show how high MR is achieved for the P arrangement of $E_z$ in L and R. (b) A general picture of field-modulated layer- and spin-resolved density of states (DOS) for the case on the right. By shifting the layer-spin dependent states in energy, the LAF field in model~(\ref{Hami}) induces a band gap in BLG, and then $E_z$ drives fully spin polarization within the gap.}
\end{figure}

\section{Model and Theory}
Our model of four vdW layers in Fig.~1(a) contains the nonmagnetic conducting BLG with Bernal stacking \cite{McKosh,Neto} in the middle and two FM insulating layers located respectively on the upper and lower sides. As a typical conducting system, BLG has the advantages of highly-adjustable electron mobility \cite{McKosh,ZhaTan} and long spin relaxation length \cite{HanKaw,Suichen,XuZhu}, which can be very little affected by vdW proximity \cite{Island,YangTu} and suggest an easily-prepared platform to realize low-consumption devices. We assume that the two insulating layers are identical 2D magnets with out-of-plane magnetization (favorable to ultrahigh density data storage \cite{WelMos}), and do not contribute to the low-energy electronic bands except for proximity-induced magnetism. An effective low-energy Hamiltonian from an empirical tight-binding model (see Appendix A) reads

\begin{equation}\label{Hami}
\begin{split}
H_0=\upsilon_0(p_x\sigma_x+\xi p_y\sigma_y)-\frac{\gamma}{2}
\sum_{i=x,y}\tau_i\sigma_i+m_z\tau_zs_z
+\lambda_U\tau_z,
\end{split}
\end{equation}
where $\sigma_i$, $\tau_i$, $\bm s_i$ ($i=x,y,z$) are, respectively,
the Pauli matrices for sublattice isospin, layer isospin and real
spin. Here, $\xi=\pm1$ indicate two opposite valleys $K$, $K'$ of the
hexagonal Brillouin zone, $p_{x(y)}$ is the momentum component,
$\upsilon_0=3at/2$ is the Fermi velocity in graphene ($a$, $t$ are
the bond length and nearest-neighbor hopping, respectively),
and $\gamma$, $m_z$ and $\lambda_U$ denote the strengths of the
interactions. The first and second terms are, respectively, the massless Dirac term
and the interlayer nearest-neighbor coupling
(usually sufficient to capture the band features
\cite{McKosh,Neto}). The third term indicates the LAF field induced by short-range magnetic proximity from two FM layers \cite{TangZhang,CarSor}. The last term represents the vertical bias.

We also consider other possible interactions which might be present but do not qualitatively change the MR effect we concern with. They read \cite{KarpCum,Island,YangTu} $H'=H'_\Delta+H'_{\rm R}$, where $H'_\Delta=\Delta\sigma_z$ represents the
intralayer inversion-symmetry breaking induced by interface lattice mismatch
and $H'_{\rm R}=\lambda_{\rm R}(\sigma_xs_y-\xi\sigma_ys_x)/2$ \cite{KarpCum,ZhaiJin1,ZhaiJin2} is the proximity-induced Rashba interaction. For simplicity, we take $H'$ the same in both monolayers. Disregarding $H'_{\rm R}$, the low-energy four bands are solved as
\begin{equation}\label{band}
\begin{split}
E^2=&\upsilon_0^2p^2+\frac{\gamma^2}{2}+\alpha^2+\Delta^2\\
&-\frac{1}{2}\sqrt{4\upsilon_0^2p^2(4\alpha^2+\gamma^2)+\left(\gamma^2-4\alpha
\Delta\right)^2},
\end{split}
\end{equation}
where $\alpha=\lambda_U+sm_z$ ($s=\pm1$ denote spin up and spin down, respectively).
In the actual numerical calculations below (Fig.~2), weak $H'_{\rm R}$ is taken into account as well.

As sketched in Fig.~1(a), the system is divided into three regions, including the left (L), right (R) terminals and the central (C) scattering region. We use
an effective empirical potential \cite{SanPra} $\lambda_U(x)=U~{\rm erf}(x/d)$ and $U$, respectively, to simulate the AP and P bias configurations. Here, erf is the error function, and $x=0$ is the center of region C.

Depending on the sample quality, the system can be a pristine LAF crystal when no DW is present in region C ($dm_z/dx=0$). The simplest DW is collinear, as sketched in Fig.~1(a), and can be simulated by a smooth potential \cite{WahAug}
\begin{equation}\label{m_z}
m_z(x)=M_0\tanh\frac{\pi x}{d}.
\end{equation}
Here, the magnetization $m_z$ vanishes at $x=0$ due to
the destructive overlap between opposite spins. We can further consider
the noncollinear Bloch
($m_y^2=M_0^2-m_z^2$,
$m_x=0$) and N\'{e}el ($m_x^2=M_0^2-m_z^2$,
$m_y=0$) types of DW due to spin fluctuation \cite{WahAug} by adding in-plane magnetization terms $m_y\tau_zs_y$ and $m_x\tau_zs_x$ to Eq.~(\ref{Hami}), respectively. The in-plane magnetization of Bloch (N\'{e}el) type is parallel (perpendicular) to the DW,
and the sign of $m_{x}$ or $m_y$ indicates the DW chirality. The spin valve functions qualitatively the same with or without DWs, however, the details of scattering potential differ.

\begin{figure*}
\centerline{\includegraphics[width=17.5cm]{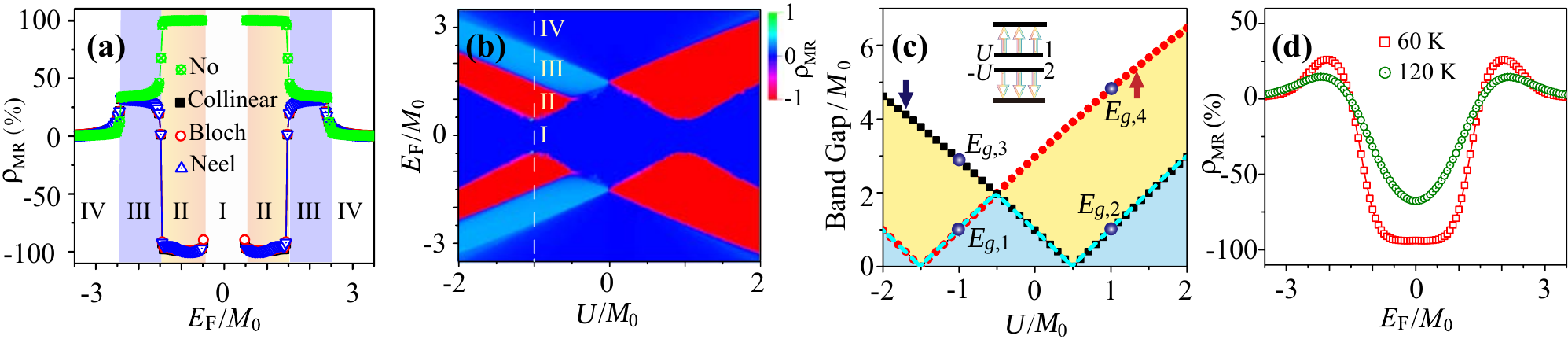}} \caption{
Calculated results from the lattice model (see Appendix A) with parameters $t=2.7$~eV, $\gamma=0.39$~eV \cite{Neto}, $M_0=27$~meV, $\Delta=13.5$~meV and $\lambda_{\rm R}=0.9$~meV. (a) MR ratio versus $E_{\rm F}$ for different DW types (no, collinear, Bloch and N\'{e}el)
for an armchair-edged strip with $W=40\sqrt3 a$ under $U=-M_0$ and $T=0$. The symbols I to IV label four MR states with clear boundaries. (b) Contour plot of MR in the $(E_{\rm F},U)$ plane for the armchair-edged strip with collinear DW at $T=0$. The dashed line corresponds to the cross-section $U/M_0 = -1$ used in (a). (c) Band gap $E_g$ in terminal L (green dashed line). For comparison, the spin-dependent band gap $E_g^s$ is shown under $\lambda_{\rm R}=0$ (red circle for spin up, black rectangle for spin down). Blue circle points denote four gap values $E_{g,i}$ ($i=1$ to 4) under $U/M_0=\pm1$. (d) Influence of temperature on MR versus $E_{\rm F}$ under $U=-M_0$ for the collinear-DW case in (a).}
\end{figure*}

Taking the collinear DW in Fig.~1(a) for example, we illustrate a typical band-to-band
tunneling to realize the high MR state for the P bias configuration due to spin mismatch
between L and R. By tuning the direction of vertical bias, the LAF spin valve can be
switched on or off. The giant MR effect is understood from a general picture of phase transition in Fig.~1(b), seen from model~(\ref{Hami}), that the LAF field drives BLG to be insulating by moving the states with opposite spins in the same layer and the states with the same spin in opposite layers oppositely. The vertical bias induces a full spin polarization by moving the states in the opposite layers oppositely. More crucial for the realization of the entirely electric control of the spin valve effect is that the opposite bias induces the opposite spin polarization. We characterize the MR effect
by a resistance ratio
\cite{LinYang,Grunberg}\label{MR}
\begin{equation}
\rho_{\rm MR}=\frac{R_{\rm AP}-R_{\rm P}}{R_{\rm AP}+R_{\rm P}},
\end{equation}
where $R_{\rm P(AP)}=1/G_{\rm P(AP)}$ is the resistance for the P
(AP) case. The conductance can be calculated by using the nonequilibrium
Green's function method \cite{Datta} as
\begin{equation}\label{conductance}
G=\frac{e^2}{h}\int_{-\infty}^{+\infty}dE{\cal
T}(E) \left(-\frac{\partial}{\partial E}f(E-E_{\rm F})\right),
\end{equation}
with the transmission given by ${\cal T}(E)={\rm Tr}({\cal
G}\Gamma_{\rm L}{\cal G}^\dag\Gamma_{\rm R})$, where $\cal G$ is the
Green's function of the system, Tr denotes the trace, $\Gamma_{\rm
L, R}=i(\Sigma_{\rm L, R}-\Sigma_{\rm L, R}^\dag)$ represents the
broadening of L and R, and $f(E-E_{\rm F})$ is the
Fermi-Dirac distribution function.

\section{RESULTS AND DISCUSSION}
We first solve the model by performing calculations in an armchair-edged sample with the width $W=40\sqrt3 a$. Two terminals L and R are
assumed to be semi-infinite. A relatively-short scattering region with the length $d=10a$
is fixed, and the main electrically-tunable parameters are the bias voltage $U$ and the Fermi energy $E_{\rm F}$ (see details in Appendix B) and Ref.~[\onlinecite{GavLaz}]). We keep $\vert U \vert$ the same for P and AP configurations and fix the parameters $t=2.7$~eV and $\gamma=0.39$~eV from bare GBL \cite{Neto} and $M_0=27$~meV, $\Delta=13.5$~meV and $\lambda_{\rm R}=0.9$~meV.

When $U=-M_0=-27$~meV is fixed ($E_z\simeq-0.5$~V/nm \cite{ZhaTan}), our results of MR ratio versus $E_{\rm F}$ in Fig.~2(a) show that nearly-perfect MR effect, {\it i.e.} $|\rho_{\rm R}|\simeq100\%$, appears for $|E_{\rm F}|/M_0\in(0.5,1.5)$, independent of whether the DW is present or not. The MR ratio of this spin valve effect changes the sign if a DW appears. This is because the $m_z$-induced spin polarization near $E_{\rm F}$ changes sign in terminal R (is unchanged in L). Nontrivially, we find the giant MR effect is not sensitive to whether the DW type is collinear, Bloch or N\'{e}el, and meanwhile, the spin valve effect is irrelevant to the magnetic chirality of DW, confirmed by changing the sign of $m_x$ or $m_y$ in calculations. To illustrate the MR effect in more detail, we show the contour plot of MR ratio in the $(E_{\rm F},U)$ plane for the collinear DW in Fig.~2(b). The result shows that there is a sharp transition between the absence of the MR effect for lower $E_{\rm F}$ and the nearly-perfect MR effect for higher $E_{\rm F}$ [see also $|E_{\rm F}|/M_0<0.5$ in Fig.~2(a)]. This parameter regime of no MR, $G_{\rm P(AP)}\equiv0$, is attributed to either of the following two factors: One is the absence of transport states induced by band gap in two terminals, and the other is the spin mismatch for both P and AP bias configurations.

To precisely clarify the result in Fig.~2(b), we plot the band gap $E_g$ (green dashed line) for terminal L as a function of $U$ in Fig.~2(c).
For contrast, we also plot the spin-dependent band gap $E_g^s$ (circle for spin up, rectangle for spin down) by ignoring $\lambda_{\rm R}$ in calculations. It is found that the total band gap changes very little if $\lambda_{\rm R}$ is added due to $\lambda_{\rm R}\ll M_0,\Delta$.
In the presence of the DW in Eq.~(\ref{m_z}), the band gap in terminal R
can be directly obtained by interchanging the tabs for opposite spins in Fig.~2(c) due to the opposite magnetic orientation in two terminals. Thus, it is sufficient to get all the information of band gap by analyzing either of two terminals. After sorting four values $E_g^{s=\pm}(\pm U)$ for terminal L in order from lowest to highest, labeled as $E_{g,i}$ ($i$=1 to 4), we find considering the tunnelling mechanism in Fig.~2(b) that MR vanishes (I, $\rho_{\rm MR}=0$) for $|E_{\rm F}|<E_{g,2}/2$, and the nearly-perfect MR (II, $|\rho_{\rm MR}|\geq95\%$) occurs under the conditions $|E_{\rm F}|\in(E_{g,2},E_{g,3})/2$ for $U<0$ and $|E_{\rm F}|\in(E_{g,2},E_{g,4})/2$ for $U>0$. Likewise, the normal giant MR (III, $95\%>|\rho_{\rm MR}|\geq 25\%$ \cite{Grunberg}) holds for $|E_{\rm F}|\in(E_{g,3},E_{g,4})/2$, and weak MR (IV, $0<|\rho_{\rm MR}|<25\%$) happens for $|E_{\rm F}|>E_{g,4}/2$.

Without a DW, the dependence of $E_g(U)$ is identical for the two terminals, the condition for MR to vanish in Fig.~2(a) is strictly given by $|E_{\rm F}|<E_g(U)/2$ [see the blue region in Fig.~2(c)]. Independent of whether the DW is present or not, the nearly-perfect MR can work for the overlapping band gap region of opposite spins [see the yellow region in Fig.~2(c)] by tuning the magnitude of the bias which can be different for L and R, in addition to the sign of the bias.

\begin{figure*}
\centerline{\includegraphics[width=13cm]{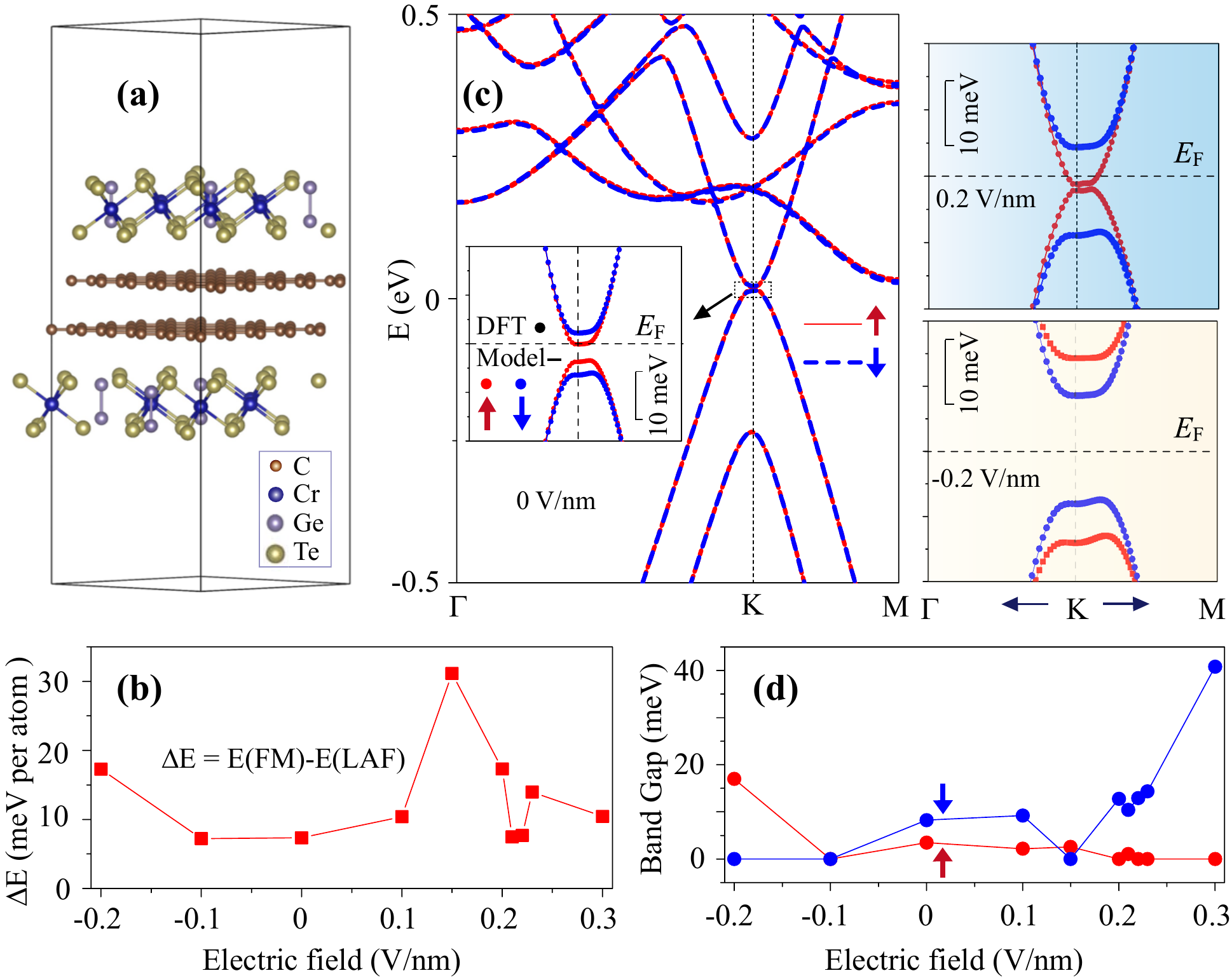}} \caption{DFT results for
CGT/BLG/CGT. (a) Side view of the optimized unit cell.
(b) The vertical electric field $E_z$ tuned energy difference on average per atom between FM and LAF states. (c) Energy bands for $E_z=0$ (left), $\pm0.2$ V/nm (right).
(d) Spin-dependent band gap as a function of $E_z$.}
\end{figure*}

Figure~2(d) further considers the influence of temperature, showing MR at $T=60$~K and 120~K, for the collinear DW at $U=-M_0$. Unsurprisingly, MR ratio versus $E_{\rm F}$ becomes smooth due to the presence of the Fermi-Dirac distribution in Eq.~(\ref{conductance}), in comparison with that in Fig.~2(a). The energy broadening induced by temperature can cover the whole band gap region, and thus enables giant MR effect when $E_{\rm F}$ lies inside the band gap. Note that the pattern of the curves in Fig.~2(d) remains the same even if the $T$-modified $M_0$ is considered in a specific system. This means, the proposed spin-valve mechanism remains operational up to the Curie temperature of the FM layers.

In particular, the numerical result $E_g^s$ for the finite-size strip in Fig.~2(c) agrees with the analytical result for the 2D bulk case $E_g^s(U)\simeq2(\gamma^2/2+\alpha^2+\Delta^2-|\gamma^2-4\alpha\Delta|/2)^{1/2}
=2|U+sm_z+\Delta|$ derived from Eq.~(\ref{band}) for $\gamma^2>4\alpha\Delta$. This reveals a negligible edge effect. Further calculations indicate that the edge effect on band gap can be ignored when $W$ exceeds a critical value $20\sqrt3a$ ($\sim$5~nm), consistent with the experimental report~\cite{YinJiang} for bare BLG. Additionally, it is argued that the electron mean free path can be estimated by $\ell\simeq\lambda_{\rm F}=h\upsilon_0/\sqrt{|E|\gamma}$ \cite{SanPra} ($\lambda_{\rm F}$ is the Fermi wavelength, $\ell\simeq248a=35.2$~nm for $E=0.01t$). The spin diffusion length may be estimated by $\ell_s^{-1}\simeq2\ell[\lambda_{\rm R}/(\hbar\upsilon_{\rm F})]^2$ (the Fermi velocity $\upsilon_{\rm F}=2\upsilon_0\sqrt{|E|/\gamma}$) in the Dyakonov-Perel mechanism \cite{DyaPerel}. The ballistic spin transport, on which we deliberately focus here, should always dominate for $d\ll\ell,\ell_s$ (see Appendix C, $\ell_s$ is usually on the ${\rm \mu m}$ scale). The nanoscale $W$ and $d$ here should be experimentally available \cite{YinJiang,WahAug}.

For zigzag-edged systems, the LAF field or vertical bias can
induce edge states \cite{ZhaiBlanter}, which reduce the MR ratio. However,
edge disorder or defects can increase the MR ratio by hindering the edge
transport. For both the armchair or zigzag-edged systems with $W<20\sqrt 3a$, the spin
valve effect can still hold in spite of strong quantum confinement, which only
leads to some quantitative variations in MR.

\section{Proposal for materials realization}
We now propose, based on DFT calculations \cite{KreFur,KreFurth,KreHaf,PerBur,TangSan}, that BLG encapsulated by CGT monolayers, each of which is a FM semiconductor \cite{GongLi,GoZha}, is feasible to realize the antiferromagnetism in our model.
There was a previous LAF prediction in CrI$_3$/BLG/CrI$_3$ \cite{CarSor}, but the formation of mixed low-energy bands from both BLG and CrI$_3$ strongly limits its superiority. For the compounds of graphene and CGT, the influence of spin-orbit couplings on low-energy bands are much smaller than other parameters and can be disregarded, as proved previously \cite{ZolGmi,ZolGmi20}.

In Fig.~3, we show the DFT results for CGT/BLG/CGT with
an average equilibrium interlayer spacing $d=3.65~{\AA}$ between CGT monolayer
and its close-contacted graphene monolayer. The unit cell in Fig.~3(a) is constructed with a $5\times5$ supercell of 50 carbon atoms per graphene layer and a $\sqrt3\times\sqrt3$ supercell of 12 chromium atoms per CGT layer. See Appendix D for other computable details. To clarify the magnetism of the ground state, we show the energy difference on average per atom ($\Delta E$) between the FM and LAF states in Fig.~3(b), which strongly suggests that the system has a LAF ground state robust against small $E_z$ that we focus on. In contrast, bilayer CGT itself is FM \cite{GongLi}, which becomes LAF
here when the layers are spatially separated.

We show the DFT band at $E_z=0$ in the left panel of Fig.~3(c), where the inset enlarges the local band near $E_{\rm F}$ and is well fitted by $\Delta=8.9$~meV, $m_z=-7.2$~meV, $t=2.441$~eV, $\gamma=0.396$~eV in our model and an additional interlayer next-nearest neighboring hopping $\gamma'=0.267$~eV.
Here, we have $m_z<0$ and $\Delta>|m_z|$, different from the assumptions in Fig.~2.
As expected, we show the contrasting spin polarization of the bands for $E_z=\pm0.2$~V/nm, which is much smaller than 1$\sim$3~V/nm experimentally accessible in vdW systems of BLG \cite{ZhaTan} and bilayer CrI$_3$ \cite{JiangLi}. We plot the spin-dependent band gap $E_g^s$ versus $E_z$ in Fig.~3(d), where spin-up and spin-down band gaps vanish for $E_z>0.23$~V/nm and $E_z<-0.1$~V/nm, respectively. This is beyond our model because CGT has a nonnegligible contribution to $E_g^s$ as $E_z$ increases.

\section{Conclusion}
Our proposal of antiferromagnetic spin valve based on
vdW magnetic proximity effect paves the way for engineering spintronic devices with low energy consumption by making full use of magneto-electric coupling. The appearance of nearly-perfect MR and the robustness of antiferromagneism against required vertical bias have revealed the superiority of our proposed device. Other adjustable quantities, {\it e.g.} stacking order, interlayer sliding, rotation angle or layer number, may bring rich opportunities for developing 2D antiferromagnetic spintronics.

\section*{ACKNOWLEDGMENTS}
This work was supported by the NSFC with Grant Nos.~12074193, 61874057 and 11874059,
the QingLan Project of Jiangsu Province (2019), Key Research Program of Frontier Sciences, CAS, Grant No. ZDBS-LY-7021, Beijing National Laboratory for Condensed Matter Physics, Natural Science Foundation of Zhejiang Province (Grant No. LR19A040002). We are grateful to S. Jiang for insightful experimental discussion.

\end{document}